\documentclass[referee]{aa}
\usepackage{graphicx}
\begin{document}

   \thesaurus{9 (02.13.2; 02.16.1;02.23.1;06.03.2)} 

   \title{A Weakly Nonlinear Alfv\'enic Pulse in a Transversely Inhomogeneous
Medium}

   \author{D. Tsiklauri, T.D. Arber and V.M. Nakariakov}

   \offprints{D. Tsiklauri (tsikd@astro.warwick.ac.uk)}

   \institute{Physics Department, University of Warwick, Coventry, 
   CV4 7AL, England}

   \date{\today}
\titlerunning{Phase mixing ...}
\authorrunning{Tsiklauri, Arber \& Nakariakov}
   \maketitle

   \begin{abstract}
The interaction of a weakly nonlinear Alfv\'enic pulse with an Alfv\'en
speed inhomogeneity
in the direction perpendicular to the magnetic field is investigated.
Identically to the phase mixing
experienced by a harmonic Alfv\'en wave, sharp transverse gradients are
generated in the pulse
by the inhomogeneity. In the initial stage of the evolution of an
initially plane Alfv\'enic
pulse, the transverse gradients efficiently generate transversely
propagating fast magnetoacoustic
waves. However, high resolution full MHD numerical 
simulations of the developed stage of the pulse
evolution show that the generation saturates due to 
destructive wave interference.
It is shown that the weakly non-linear description of the
generated fast magnetoacoustic wave is well described by
the driven wave equation proposed in Nakariakov, Roberts \&
Murawski (1997), and a simple numerical code
(2D MacCromack), which solves it with minimal CPU resources
produces  identical results to those obtained from
the full MHD code ({\it Lare2d}, Arber et al. 2001).
A parametric study of the phenomenon is undertaken, showing that,
contrary to one's
expectations, steeper inhomogenities of the Alfv\'en speed do not
produce higher saturation levels
of the fast wave generation. There is a certain optimal gradient of the
inhomogeneity that
ensures the maximal efficiency of the fast wave generation.

\end{abstract}

\section{Introduction}

Many astrophysical objects and space plasma systems 
are significantly influenced by magnetohydrodynamic (MHD) waves. 
It is enough to mention that MHD waves constitute the basis
of many current theories  of solar and stellar
wind acceleration, the heating of solar and stellar coronae
and the stability
of molecular clouds, etc.
In solar physics interest in MHD waves has increased
significantly with the successful detection of 
MHD waves in coronal loops and plumes by recent
observations from the SOHO and TRACE spacecrafts.
This, in turn, has generated lively interest in MHD
waves in the theoretical solar physics community
in the context of coronal plasma heating and solar wind acceleration
(see Roberts 2000 for a review).

A significant aspect in MHD wave dynamics, which
has attracted considerable attention recently 
is the interaction of waves with
{\it plasma inhomogeneities}. Because of the essentially anisotropic nature
of MHD waves, they are affected by the 
inhomogeneities in many ways. In particular, structuring of the medium
leads to the appearance of long wave length dispersion, the existence of local
resonances associated with e.g. resonant absorption, and Alfv\'en wave 
phase mixing
(see, e.g. Roberts 1991 and Roberts \& Ulmschneider 1997). 

In a majority of astrophysical applications, almost incompressible 
Alfv\'en waves
are subject to very weak dissipation. Therefore, physical mechanisms which
can create small scales in the waves, enhancing the dissipation, 
are very important.
One of these mechanisms is phase mixing of Alfv\'en waves 
interacting with a plasma
inhomogeneity. When an initially plane Alfv\'en wave
propagates along a straight magnetic field, the presence of 
an Alfv\'en speed inhomogeneity
across the field leads to distortion of the wave front. In other words, each
magnetic interface supports Alfv\'en waves propagating with its own local
Alfv\'en speed. After a while, the neighbouring magnetic interfaces are oscillating
out of phase and the Alfv\'en wave becomes "phase-mixed". 
This phenomenon decreases dramatically the perpendicular wave length and generate
very sharp transverse gradients. In the presence of finite  viscosity or resistivity,
the short wavelengths are subject to enhanced dissipation. Since this mechanism
was proposed by Heyvaerts \& Priest (1983) for heating the solar corona,
the linear regime of Alfv\'en wave phase mixing
has been intensively investigated by many researchers, mainly in the coronal heating
context (Abdelatif 1987, Browning 1991, Ireland \& Priest 1997, see also the 
critical paper of Parker 1991), 
including investigation of Alfv\'en wave phase mixing in inhomogeneous steady flows
(Ryutova \& Habbal 1995, Nakariakov, Roberts \& Murawski 1998) and in two dimensional
structures (Hood, Ireland \& Priest 1997, Ruderman, Nakariakov, \& Roberts 1998, 
Ruderman et al. 1999,
De Moortel et al. 1999, De Moortel, Hood \& Arber 2000) and studies of secondary Kelvin--Helmholtz
instabilities (Browning \& Priest 1984). 
Possible manifestation of this effect in coronal observations have been  discussed (Ireland 1996, 
De Moortel \& Hood 2000) and numerical simulations of Alfv\'en wave phase mixing 
have been
undertaken in various astrophysical and geophysical contexts 
(Botha et al. 2000, Grappin, Leorat \& Buttighoffer 2000, 
Malara, Primavera \& Veltri 1996,
Ofman \& Davila 1997).

In the linear regime of Alfv\'en wave phase mixing, the wave does not interact with
other wave modes, and the wave dissipation is enhanced 
due to the generation of small scales across the field. In the nonlinear regime,
Alfv\'en waves interact with, in particular, magnetoacoustic waves.
For the Alfv\'en wave amplitudes expected the lower corona of the Sun, the linear treatment of the
process of phase mixing seemed to be justified (e.g. Sakurai \& Granik 1984) as the Alfv\'en
waves are subject to cubic, not quadratic as for magnetoacoustic waves, nonlinearity.
The presence of structuring can dramatically affect the efficiency of nonlinear processes.
In coronal holes, outwardly propagating Alfv\'en waves are amplified by stratification and,
therefore, quickly reach the nonlinear regime (Ofman \& Davila 1997, 
Nakariakov, Ofman \& Arber 2000).
In the presence of structuring across the field, phase mixing leads to nonlinear generation
of transversely propagating fast magnetoacoustic waves (Nakariakov, Roberts \& Murawski 1997, 1998).

The nonlinear generation of fast waves, based upon a combination of inhomogeneous and nonlinear
effects, can provide new physical mechanisms for MHD wave dissipation.
In the linear regime of phase mixing, Alfv\'en waves dissipate due to 
the {\it shear} component of
the viscosity tensor, while the fast magnetoacoustic 
waves dissipate due to the {\it bulk} viscosity. 
In the nearly collisionless coronal
plasma the bulk viscosity is estimated to be $10^{10}$ times larger than the
shear viscosity. Thus, fast magnetoacoustic waves that are
generated by Alfv\'en waves dissipate much more
efficiently than the Alfv\'en waves.  This mechanism
can be regarded as the indirect heating of plasma
by phase mixing.

In the corona of the Sun there is some evidence that the shear viscosity
may actually be  
of the same order as the bulk viscosity
(Nakariakov et al. 1999). In this case, the mechanism of indirect 
heating is still very important,
because it provides a possibily of heat distribution 
across the magnetic field.
Indeed, the transversely propagating fast waves carry the energy across the field and deposit it at
some distance from the region of phase mixing. Regular thermal conduction is depressed in the
perpendicular direction by the magnetic field. Thus, the mechanism of indirect heating
of astrophysical plasmas by Alfv\'en wave phase mixing requires serious investigation.

Nakariakov, Roberts \& Murawski (1997), by applying a weakly nonlinear description
for the dynamics of the generated fast magnetoacoustic waves, pointed out that phase mixed
Alfv\'en waves had to generate secularly growing fast waves (their Eq.~(28)). This 
analytical result was successfully reproduced in full-MHD numerical simulations of the
{\it initial} stage of the fast wave generation.
Botha et al. (2000) undertook numerical simulations of this phenomenon over much longer
times, investigating the efficiency of
the indirect heating by considering {\it harmonic} phase-mixed 
Alfv\'en perturbations. They
found that initially the amplitude of the generated fast magnetoacoustic waves 
grows but then saturates and does not 
reach a substantial fraction of the Alfv\'en wave amplitude. They 
further postulated that
this saturation was due to the nature of the source term for fast waves. For harmonic 
Alfv\'en waves, as distinct from the single pulse considered in this paper, 
phase mixing
leads to there being many fast wave sources in the direction of the background
inhomogeneity. Indeed the number of such sources, and their spacing, changes in time
and as a consequence the fast wave sources cannot remain coherent and destructive
wave interference limits the growth of fast wave energy.

In this work, we simulate numerically the interaction of a linear Alfv\'enic pulse 
with a one-dimensional, perpendicular to the magnetic field, plasma inhomogeneity. 
The pulse phasemixes leading to steep transverse gradients.
This leads to the generation of fast magnetoacoustic waves.
This work has three objectives: to investigate 
whether a phase-mixed  Alfv\'enic pulse,
as opposed to Botha et al. 2000's harmonic Alfv\'en wave,
can efficiently generate fast magnetoacoustic waves;
to check whether the rapid growth (in time)
of the generated fast magnetoacoustic waves still holds
at the later stages and to clarify the role of the destructive wave interference
for an isolated pulse.

The paper is organized as follows: in section 2 we outline the 
basic aspects of our model.
Section 3 presents the results of our numerical calculation.
While in section 4 we close with a brief discussion of the
main results of this paper.

\section{The model}

In our model we use equations of cold (zero plasma-$\beta$)
ideal MHD
$$
\rho {{\partial \vec V}\over{\partial t}} + 
\rho(\vec V \cdot \nabla) \vec V = -{{1}\over{4 \pi}}
\vec B \times {\rm curl} \vec B, \eqno(1)
$$
$$
{{\partial \vec B}\over{\partial t}}= {\rm curl} (\vec V \times \vec B),
\eqno(2)
$$
$$
{\rm div} \vec B =0, \eqno(3)
$$
$$
{{\partial \rho}\over{\partial t}} + {\rm div}(\rho \vec V)=0,
\eqno(4)
$$
where $\vec B$ is the magnetic field, $\vec V$ is plasma velocity,
and $\rho$ is plasma mass density.

We consider equations (1)-(4) in Cartesian coordinates ($x,y,z$)
and for simplicity assume that there is no variation in the
$y$-direction, i.e. ($\partial / \partial y =0$). The equilibrium
state is taken to be an inhomogeneous plasma of density $\rho_0(x)$
and a uniform magnetic field
$B_0$ in the $z$-direction.

Nakariakov, Roberts \& Murawski 1997 have obtained a set of governing equations
for the finite amplitude perturbations for the above formulated
physical system (see, also, Nocera, Priest \& Hollweg 1986), 
and for brevity we do not repeat their derivation
here. However, we quote their final analytic result, which is
two coupled, non-linear wave equations describing
non-linear coupling of Alfv\'en and fast magnetoacoustic waves
$$
{{\partial^2  V_y}\over{\partial t^2}} -
C_A^2(x){{\partial^2  V_y}\over{\partial z^2}}
={{1}\over{\rho_0(x)}}\left( {{\partial N_2}\over{\partial t}}
+{{B_0}\over{4 \pi}}{{\partial N_5}\over{\partial z}} \right),
\eqno(5)
$$
$$
{{\partial^2  V_x}\over{\partial t^2}} -
C_A^2(x)\left({{\partial^2  V_x}\over{\partial x^2}}+
{{\partial^2  V_x}\over{\partial z^2}}\right)
={{1}\over{\rho_0(x)}}\left( {{\partial N_1}\over{\partial t}}
+{{B_0}\over{4 \pi}}{{\partial N_4}\over{\partial z}} -
{{B_0}\over{4 \pi}}{{\partial N_6}\over{\partial x}}\right),
\eqno(6)
$$
where $C_A(x)=B_0/\sqrt{4 \pi \rho_0(x)}$ is a local Alfv\'en
speed and following notation has been used
$$
N_1=-\rho{{\partial V_x}\over{\partial t}}-(\rho_0(x)+\rho)
\left( V_x {{\partial V_x}\over{\partial x}} + 
 V_z {{\partial V_x}\over{\partial z}}\right)
+{{B_z}\over{4 \pi}} \left({{\partial B_x}\over{\partial z}}
-   {{\partial B_z}\over{\partial x}}\right)-
{{B_y}\over{4 \pi}} {{\partial B_y}\over{\partial x}},
\eqno(7)
$$
$$
N_2=-\rho{{\partial V_y}\over{\partial t}}-(\rho_0(x)+\rho)
\left( V_x {{\partial V_y}\over{\partial x}} +
 V_z {{\partial V_y}\over{\partial z}}\right)
+{{B_x}\over{4 \pi}} {{\partial B_y}\over{\partial x}}
+ {{B_z}\over{4 \pi}} {{\partial B_y}\over{\partial z}},
\eqno(8)
$$
$$
N_4=-{{\partial}\over{\partial z}}(V_zB_x-V_xB_z),
\eqno(9)
$$
$$
N_5=-{{\partial}\over{\partial z}}(V_zB_y-V_yB_z)
+{{\partial}\over{\partial x}}(V_yB_x-V_xB_y),
\eqno(10)
$$
$$
N_6={{\partial}\over{\partial x}}(V_zB_x-V_xB_z).
\eqno(11)
$$
As noted in Nakariakov, Roberts \& Murawski 1997, the non-linear dynamics
of Alfv\'en waves is described by Eq.(5). Whereas, Eq.(6)
is the governing equation for the non-linear dynamics of
finite amplitude fast magnetoacoustic waves. In the linear limit
when the right-hand-side terms in the Eqs.(5) and (6) tend to
zero, Alfv\'en and fast magnetoacoustic waves propagate
along the applied magnetic field independently.
Therefore, in this case Eq.(5) describes phase-mixed
Alfv\'en waves, which propagate along $z$-axis perturbing
physical quantities in the $xOz$ plane. Whereas, Eq.(6)
describes either trapped or leaking fast magnetoacoustic waves 
(depending whether there is a maximum or minimum in the density 
profile $\rho_0(x)$).

We consider the physical situation when  initially fast magnetoacoustic
perturbations are absent and the  initial amplitude of
the Alfv\'en wave is small. In this case,
if the Alfv\'en wave is  initially a plane wave, the subsequent
evolution of the wave, due to the difference in local Alfv\'en speed
across the $x$-coordinate, leads to the distortion of the wave front.
Hence the appearance of transverse (with respect to the 
applied magnetic field) gradients, which grow linearly with time.
Therefore the leading non-linear term on the right-hand-side
of Eq.~(6) is the transverse gradient of the 
magnetic pressure perturbation,
i.e. to a fairly good accuracy (which is substantiated by our
numerical calculations presented below)
the dynamics of the fast magnetoacoustic waves can
be described by an approximate equation (Nakariakov, Roberts \& Murawski 1997)
$$
{{\partial^2  V_x}\over{\partial t^2}} -
C_A^2(x)\left({{\partial^2  V_x}\over{\partial x^2}}+
{{\partial^2  V_x}\over{\partial z^2}}\right)
=-{{1}\over{4 \pi \rho_0(x)}} {{\partial}\over{\partial t}}
\left(
B_y{{\partial B_y}\over{\partial x}} \right).
\eqno(12)
$$

\section{Numerical Results}

As in the work of Botha et al. 2000, 
we have used the following background density profile
$$
\rho_0(x)=3-2 \, \tanh(\lambda x) \eqno(13).
$$
Here, $\lambda$ is a free parameter which controls
the steepness  of the density profile gradient. In our 
normalization, which is the same as that of
Botha et al. 2000, the local Alfv\'en speed is
$C_A(x)=1/ \sqrt{3-2 \, \tanh(\lambda x)}$, and
we plot it in Fig.1 for the parameters used in our 
numerical simulations (see below). As can be seen
from the plot the local Alfv\'en speed has a gradient
at $x=0$ and its value is controlled by the 
$\lambda$-parameter.
\begin{figure}
     \resizebox{\hsize}{!}{\includegraphics{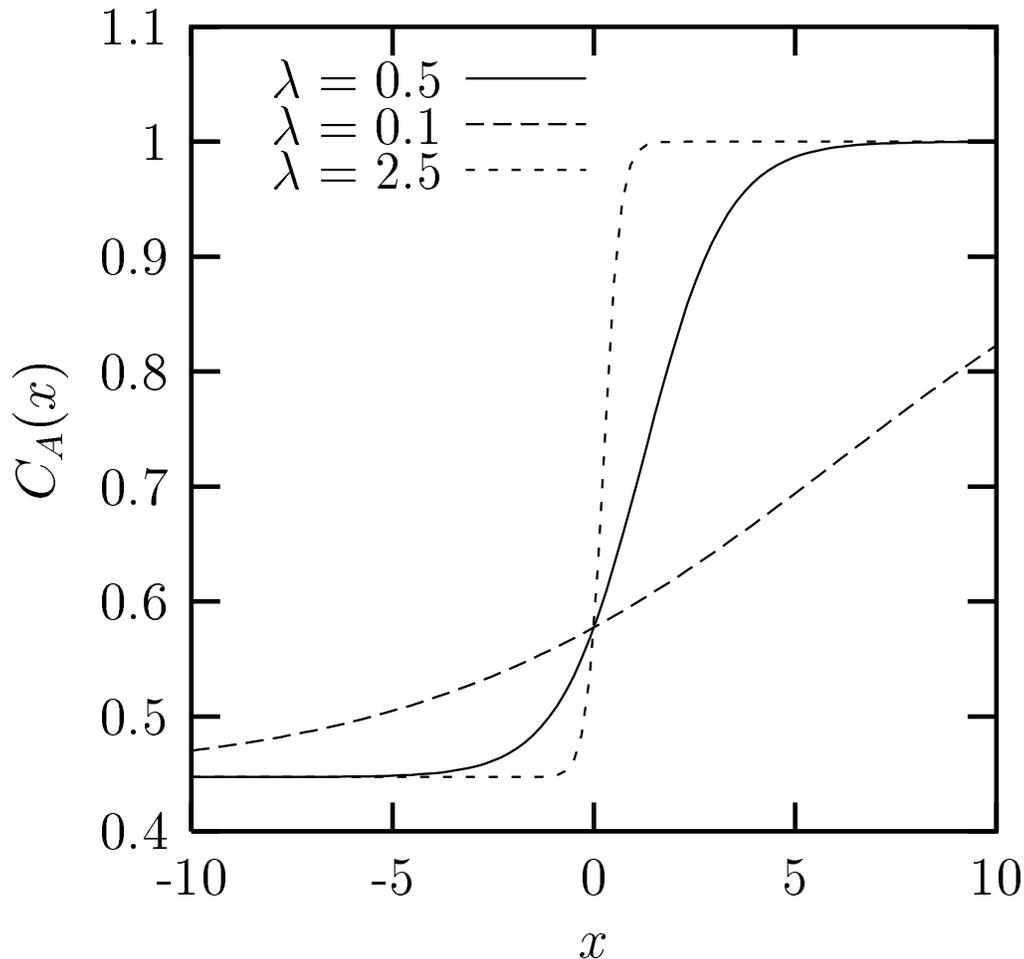}}
     \caption{Dimensionless  Alfv\'en speed $C_A(x)$ versus $x$ for
$\lambda=0.5$ solid line, $\lambda=0.1$ long dashed line, 
$\lambda=2.5$ short dashed line.}
\label{fig1}
\end{figure}
Full numerical calculation of the non-linear set of MHD
equations (1)-(4), in the above mentioned geometry,
has been performed using {\it Lare2d} (Arber et al. 2001).
{\it Lare2d} is a numerical code which operates by taking
a Lagrangian predictor-corrector time step and after each
Lagrangian step all variables are conservatively
re-mapped back onto the original Eulerian grid
using Van Leer gradient limiters.
This code was also 
used to produce the results in Botha et al. 2000.

We set up the code in such a way that initially fast magnetoacoustic
perturbations are absent and the  initial amplitude of
the Alfv\'en pulse is small, i.e. A=0.001. 
In all our numerical runs plasma $\beta$ was fixed at 0.001.
In the numerical simulations
the Alfv\'en perturbation 
is  initially a plane (with respect to $x$-coordinate) 
pulse, which has a Gaussian structure in the $z$-coordinate,
and is moving at the local Alfv\'en speed $C_A(x)$:
$$
B_y(x,z,t)=A\exp \left(-{{(z-C_A(x)\, t)^2}\over{\delta}}\right)
\eqno(14).
$$
Here, $\delta$ is a free parameter which controls the width
of the  initial Gaussian Alfv\'en pulse.
The simulation box size is set by the limits $-17.5 < x < 17.5$
and  $-27.5 < z < 27.5$ and the pulse starts to move from
point $z=-25.0$ towards the positive $z$'s.
We have performed calculation on various resolutions in
order to achieve convergence of the results.
The graphical results presented here are for the spatial
resolution $4400 \times 1400$, which refers to number of
grid points in $z$ and $x$ directions respectively.
\begin{figure}
     \resizebox{\hsize}{!}{\includegraphics{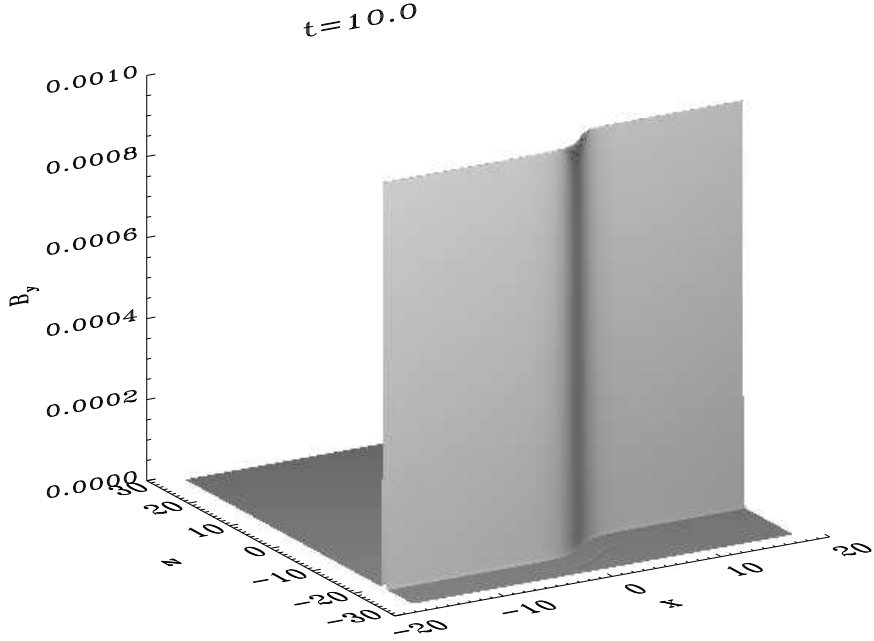}}
     \caption{Solution of the full MHD equations (1)-(4) using
Lare2d code, $B_y$, at $t=10$.}
\label{fig2}
\end{figure}
In Fig. 2 we show how the initial pulse ($B_y$) has evolved 
by $t=10$. It can be seen from the graph that 
because of the difference in local Alfv\'en speeds
(recall from Fig. 1 that for $x<0$ $C_A(x)=1/\sqrt{5}$ and 
for $x>0$ $C_A(x)=1$) the initially flat (with respect to $x$-coordinate)
Gaussian pulse has been distorted along $x=0$ axis.
This distortion of the pulse front creates a transverse
gradient ($\partial B_y / \partial x$) which according
to Eq. (12) is a driving force for the generation of
fast magnetoacoustic wave ($V_x$).
At time $t=0$ fast magnetoacoustic wave was absent from the
system. While at time $t=10$ we see from Fig. 3 that
$V_x$ has grown up to the square of  initial Alfv\'en perturbation
amplitude i.e. $10^{-6}$.

Fig. 4 presents a contour plot of $B_y$ at time $t=50$.
We observe that the Alfv\'en pulse has been distorted
even further, thus creating even stronger transverse
gradients which drive the growth of fast magnetoacoustic
waves. In Fig. 5 we plot generated $V_x$ at the same time
instance, and we observe that its maximal value has grown
further up to $1.78 \times 10^{-6}$. 
\begin{figure}
     \resizebox{\hsize}{!}{\includegraphics{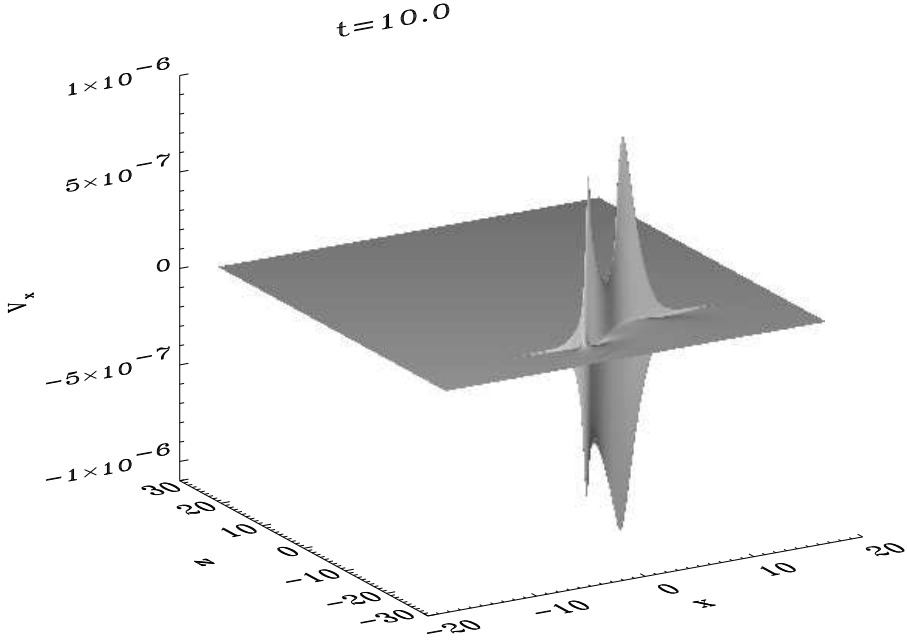}}
     \caption{Solution of the full MHD equations (1)-(4) using
Lare2d code, $V_x$, at $t=10$.}
\label{fig3}
\end{figure}

We have kept track of the maximal value of generated
fast magnetoacoustic wave, 
$V_x^a(t)=max(|V_x(x,z,t)|)$, as the simulation
progresses. The results are presented by the
think solid line in Fig. 6. As we mentioned above,
at the early stages of evolution, the dynamics of
generated fast magnetoacoustic wave is governed by Eq.(12).
In order to prove this, we have written a code
which, instead of solving the fully non-linear set of MHD equations
(1)-(4), solves the weakly non-linear Eq.(12) directly,
using a second order MacCormak scheme,
for the driver given by
Eq.(14). This driver can be easily written
by substituting Eq.(14) into Eq.(12):

$$
D(x,z,t)= -{{2A^2}\over{\delta}} C_A^2(x) {{d C_A(x)}\over{d x}}
\left[ {{4}\over{\delta}}(z-C_A(x)\, t)^2C_A(x)\, t +
(z-2C_A(x)\, t) \right] \times
$$
$$
\exp \left(-{{2(z-C_A(x)\, t)^2}
\over{\delta}}\right).
\eqno(15)
$$
where $D(x,z,t)$ is the right hand side of Eq (12).
\begin{figure}
     \resizebox{\hsize}{!}{\includegraphics{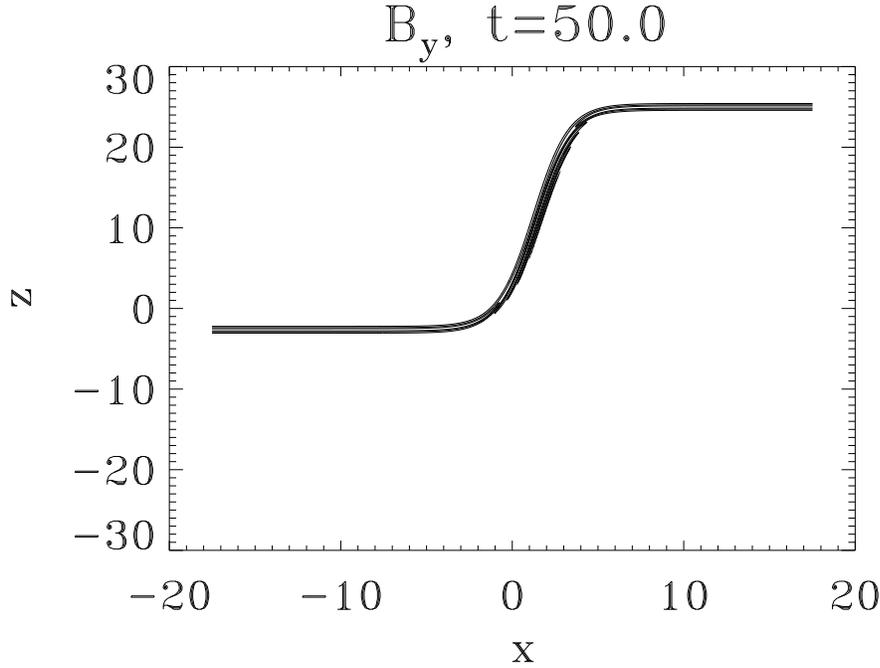}}
     \caption{Contour plot of solution of the full MHD equations (1)-(4) 
     using Lare2d code, $B_y$, at $t=50$.}
\label{fig4}
\end{figure}
We have solved Eq.(12) on a simulation box $-30 < x < 30$
and  $-30 < z < 30$ using different spatial resolution.
The results for the $V_x^a(t)$ are plotted in Fig. 6.
We observe that for $4000 \times 4000$ resolution 
(thin solid line) the behavior of the maximal value of
generated fast magnetoacoustic wave is quite similar
to the results produced by the fully non-linear {\it Lare2d}
code.
\begin{figure}
     \resizebox{\hsize}{!}{\includegraphics{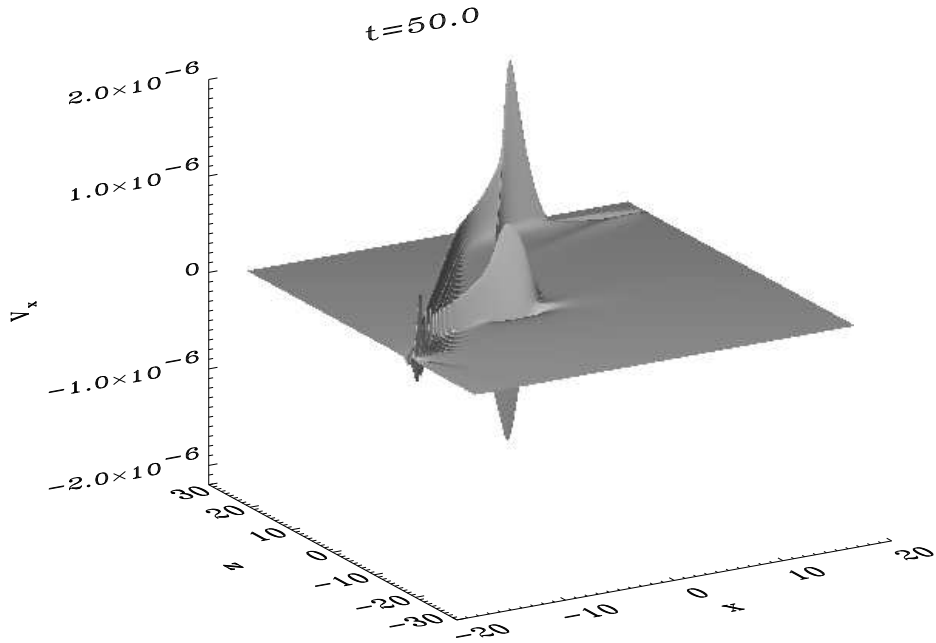}}
     \caption{Solution of the full MHD equations (1)-(4) using
Lare2d code, $V_x$, at $t=50$.}
\label{fig5}
\end{figure}
In Fig. 7 we plot generated $V_x$ at time instance
$t=50$ obtained by the 2D MacCormak code which solves Eq.(12).
The similarity with Fig. 5 is obvious, which validates
our earlier statement that Eq.(12) describes
the dynamics of the generated fast magnetoacoustic wave
to a sufficient accuracy {\it even at later times}.

As can be seen from Eq.(15) the driver of Eq.(12)
is a third order polynomial in time. Thus, one would
expect continuous secular growth of $V_x^a(t)$. However, as can be
seen from Fig. 6, $V_x^a(t)$  at the later stages
grows more slowly than initially. 
\begin{figure}
     \resizebox{\hsize}{!}{\includegraphics{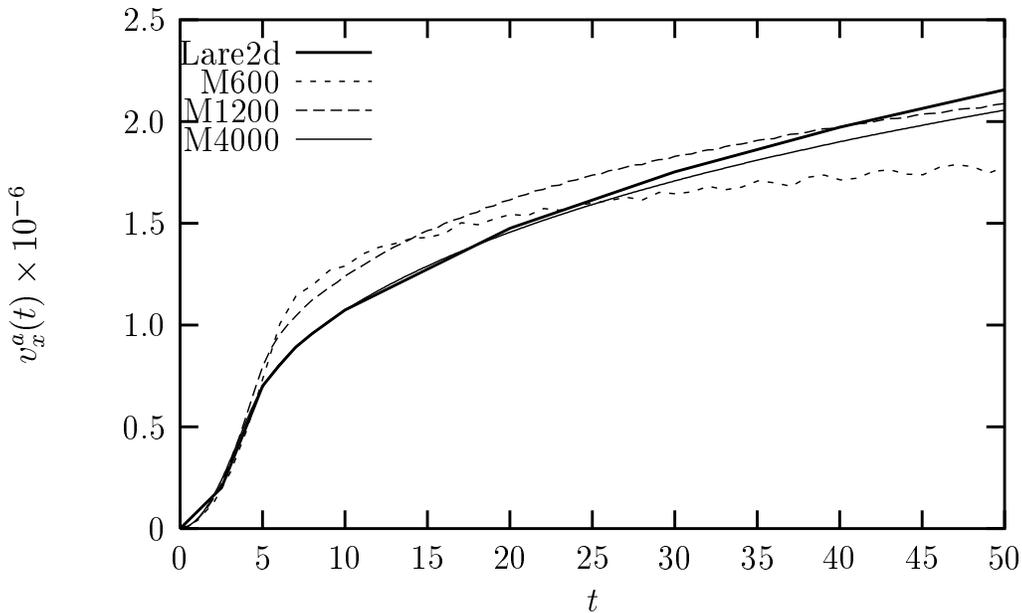}}
     \caption{Maximal value of the generated
fast magnetoacoustic wave, $V_x^a(t)$, versus time.
Thick solid curve represents the solution of the 
full MHD equations (1)-(4) using Lare2d code.
Thin solid, long dashed and short dashed curves represent
solutions of the Eq.(12) using 2D MacCormak code on
$4000 \times 4000$, $1200 \times 1200$ and $600 \times 600$
spatial resolution accordingly.}
\label{fig6}
\end{figure}
The situation is quite similar to what has been observed
by Botha et al. 2000. Namely, in spite of secularly
growing driver the solution of Eq.(12) saturates.
Botha et al. 2000 have speculated that 
the saturation of fast magnetoacoustic wave is due to the
destructive wave interference
effect. Thus, in order to bring clarity into the situation,
following Botha et al. 2000, we also did calculation of the
1D analog of Eq.(12).
We have solved equation
$$
{{\partial^2  V_x}\over{\partial t^2}} -
C_A^2(x){{\partial^2  V_x}\over{\partial x^2}}=
D_1(x,t), \eqno(16)
$$
where $D_1(x,t)=D(x,(C_A(+\infty)- C_A(-\infty))\,t, t)$. In other words
we put $z \to (C_A(+\infty)- C_A(-\infty))\,t$, 
which physically means that
we reduce 2D problem into 1D problem by
making a transformation to
 the frame which moves with the mid-point of phase mixed
Alfv\'en pulse moving with velocity $(C_A(+\infty)- C_A(-\infty))\,t$.
\begin{figure}
     \resizebox{\hsize}{!}{\includegraphics{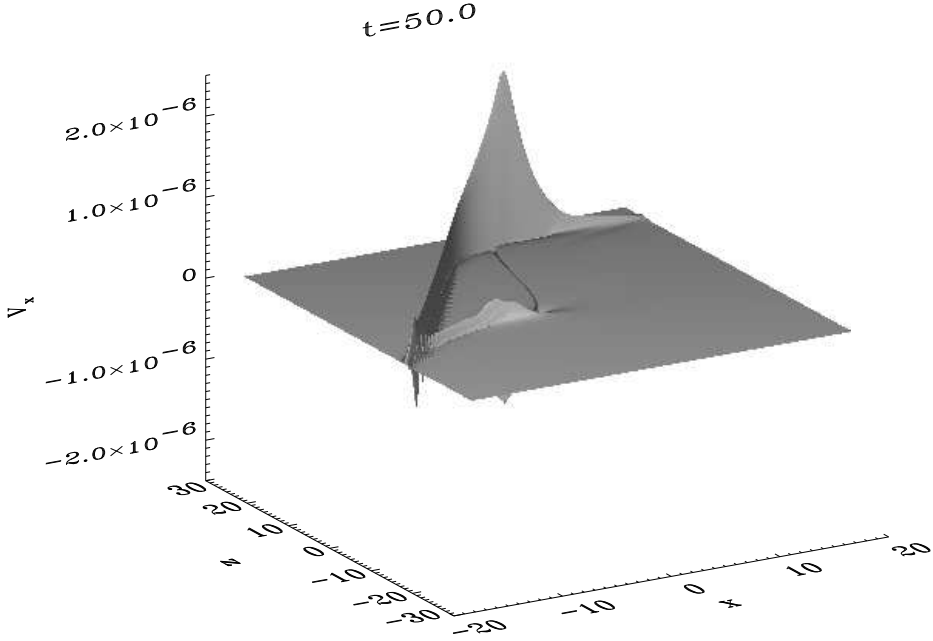}}
     \caption{Solution of the Eq.(12) using 2D MacCormak code with
$4000 \times 4000$ spatial resolution, $V_x$, at $t=50$.}
\label{fig7}
\end{figure}
We solved Eq.(16) with various spatial resolutions to
make sure that the results converge. 
In Fig. 8 we plot the solution of Eq.(16) on
the simulation interval $-55 <x <55$ using 20000
grid points. At time $t=0$ $V_x$ is zero.
Top row shows $V_x$ at time instances $t=1,10,50$.
We see that the solution comprises two pulses
outgoing into positive and negative directions
at speeds  $C_A(x)=1/\sqrt{5}$ for $x<0$ and 
$C_A(x)=1$ for $x>0$.
The bottom row contains plots of the driver, $D_1(x,t)$, 
at the same instances.
Interestingly, we observe that initially, the  driver is
asymmetric and has larger positive part, however
at the later stages it becomes symmetric and 
progressively narrow. This means that, similarly to
results obtained by Botha et al. 2000, the
fast magnetoacoustic waves generated by such a driver
will saturate because the positive and negative parts
of the driver, which get closer as time progresses,
cancell each other due to destructive wave interference. 
The difference between
our results and the ones by Botha at al. 2000
is that since we have considered a single
Gaussian pulse (hoping that single pulse will not
trigger destructive wave interference effect) we have a
single symmetric driver. Botha at al. 2000 considered
a harmonic initial Alfv\'en perturbation, thus
obtained a driver with a number of symmetric 
peaks, who's number grows in time (see Fig. 6 in
Botha et al. 2000 for comparison).
\begin{figure}
     \resizebox{\hsize}{!}{\includegraphics{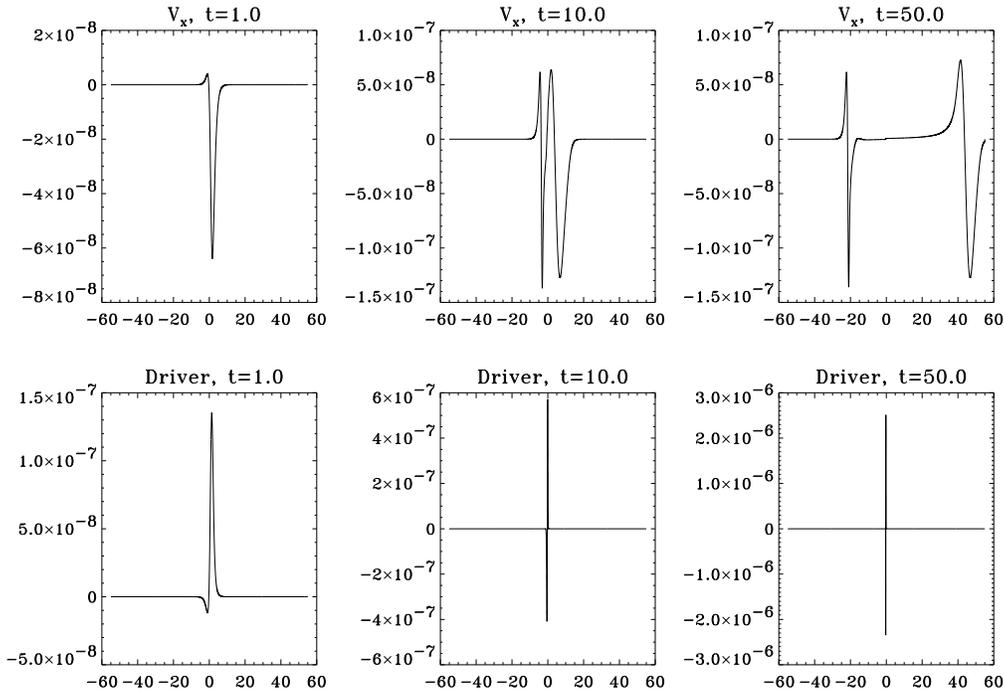}}
     \caption{Top row: solution of Eq.(16) using 1D MacCormak code on
20000 spatial resolution, $V_x$, at times $t=1,10,50$;
bottom row: driver of Eq.(16), $D_1(x,t)$, at times $t=1,10,50$.}
\label{fig8}
\end{figure}
We have kept track of the maximal value of generated
fast magnetoacoustic wave, $V_x^a(t)$, as our 1D code
progressed in time.
The results for the three values of
$\lambda$ are presented in Fig. 9.
It should be noted that there is a notable
difference between 2D  (Fig.6) and 1D (Fig. 9) results.
In 2D, as opposed to 1D,
 we do not observe clear saturation of $V_x^a(t)$.
We can only conclude form Fig. 6 that
 at the later stages $V_x^a(t)$
grows much slower than initially.
This can be explained by the fact that
in 1D, we define $V_x^a(t)$ as a maximal value of generated
fast magnetoacoustic wave in 
the frame {\it which moves with the mid-point} of phase mixed
Alfv\'en pulse having the velocity $(C_A(+\infty)- C_A(-\infty))\,t$.
Therefore in 2D the quantity that would be more
appropriate to compare to $V_x^a(t)$ from 1D results is
the maximal value of generated
fast magnetoacoustic wave in 1D slices along $x$-coordinate
which are taken at $z = (C_A(+\infty)- C_A(-\infty))\,t$
as the 2D MacCormak code progress in time.
We denote the latter quantity by 
$V_x^m(t)=max(|V_x(x,z = (C_A(+\infty)- C_A(-\infty))\,t,t)|)$.
\begin{figure}
     \resizebox{\hsize}{!}{\includegraphics{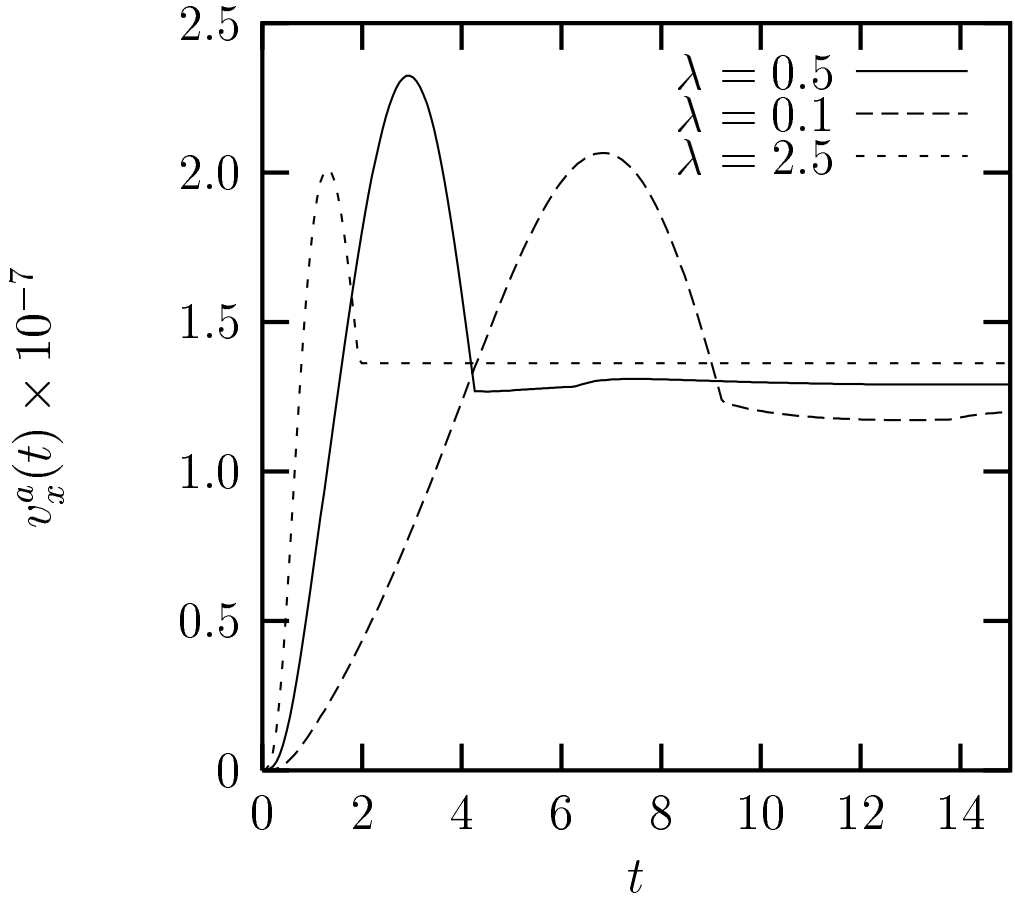}}
     \caption{Maximal value of the generated
fast magnetoacoustic wave, $V_x^a(t)$, versus time.
Solid, long dashed and short dashed curves represent
solutions of Eq.(16) using 1D MacCormak code on
200000 spatial resolution for $\lambda=0.5,0.1,2.5$ accordingly.}
\label{fig9}
\end{figure}
In Fig. 10 we plot $V_x^m(t)$ for three values
of $\lambda=0.1,0.5,2.5$. These are the same values
that are used in Fig.1. The $\lambda$-parameter controls
strength of the transversal gradient.
The noteworthy features that can be observed from
the plot are as following: stronger transverse gradients (larger
$\lambda$'s) cause earlier start of the fast magnetoacoustic waves
saturation process, and in turn, yield lower saturation levels. 
Whereas, weaker gradients (smaller
$\lambda$'s) result in later onset of the saturation, and in turn,
higher saturation levels.
This result,  actually, validates our explanation 
of the fast magnetoacoustic wave saturation by the
destructive wave interference effect. Namely, larger $\lambda$
means steeper transverse gradient, i.e. narrower driver
acting from the {\it beginning}. Therefore, destructive wave interference
effect starts to a play role earlier yielding faster shut down
in the growth of $V_x$. For the the weaker gradients, i.e.
small $\lambda$, the opposite is true.
This result is in agreements with that
obtained by Botha et al. 2000 (see Fig. 3 in
Botha et al. 2000 for comparison).
\begin{figure}
     \resizebox{\hsize}{!}{\includegraphics{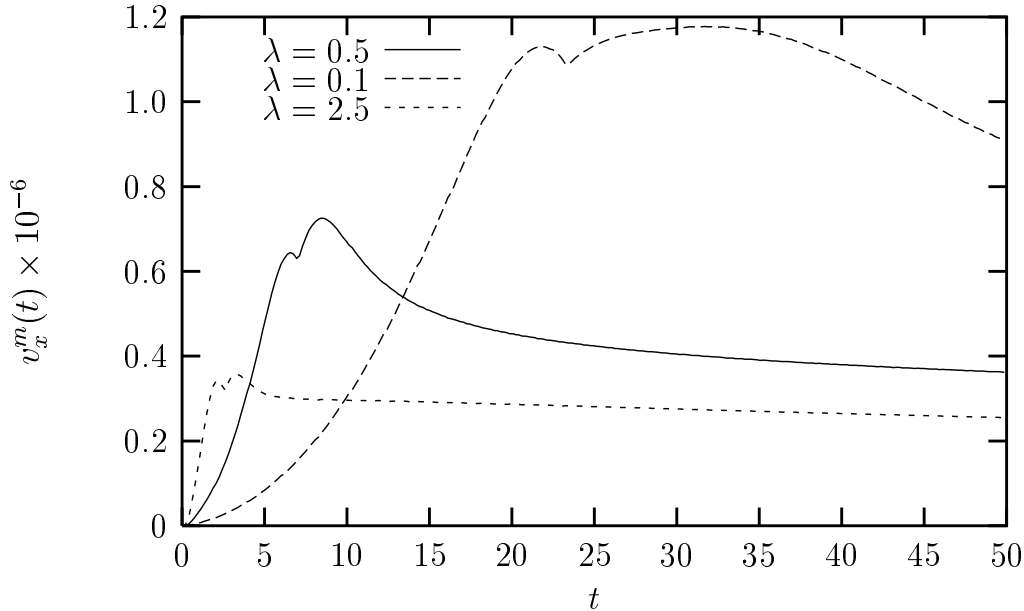}}
     \caption{Maximal value of the generated
fast magnetoacoustic wave, $V_x^m(t)$, in 1D slices along $x$-coordinate
which are taken at $z = (C_A(+\infty)- C_A(-\infty))\,t$ versus time.
Solid, long dashed and short dashed curves represent
solutions of Eq.(12) using 2D MacCormak code on
$4000 \times 4000$ spatial resolution for $\lambda=0.5,0.1,2.5$ 
accordingly.}
\label{fig10}
\end{figure}
Another, important observation from Fig.10 is that the
saturation of $V_x^m(t)$ is similar to
that from 1D results (Fig.9). This differs from the results shown in
Fig. 6 in which the amplitude of $V_x$ continues to grow albeit not
at the rapid initial phase from $t=0$ to $t=5$. The discrepancy between
the 1D and 2D results can be understood from the form of the 2D solution
in Fig. 5. Here it is clear that the maximum $V_x$ in the box is not on the
$x$ line used to produce Fig. 10 but is located at regions where
the background density levels off to a constant value. At these locations
the $x$ gradient of $B_y^2$ remains small but does not contract in $x$
as rapidly as in the centre of the phasemixing region. Thus these regions
are a source of continued secular growth in $V_x^a$ for longer.

Following Botha et al. 2000 we have performed
a study of parameter space, by investigating the
dependence of the fast magnetoacoustic wave
saturation levels upon the two free parameters
in our problem, namely, $\lambda$ and $\delta$,
using the 2D MacCormak code. In Fig. 11 we plot $V_x^a(t)$ for three values
of $\lambda=0.1,0.5,2.5$. 
Again, the noteworthy features  are: stronger transverse gradients (larger
$\lambda$'s) cause earlier start of the fast magnetoacoustic waves
saturation process, and in turn, yield lower saturation levels. 
Whereas, weaker gradients (smaller
$\lambda$'s) result in later onset of the saturation, and in turn,
higher saturation levels.
\begin{figure}
     \resizebox{\hsize}{!}{\includegraphics{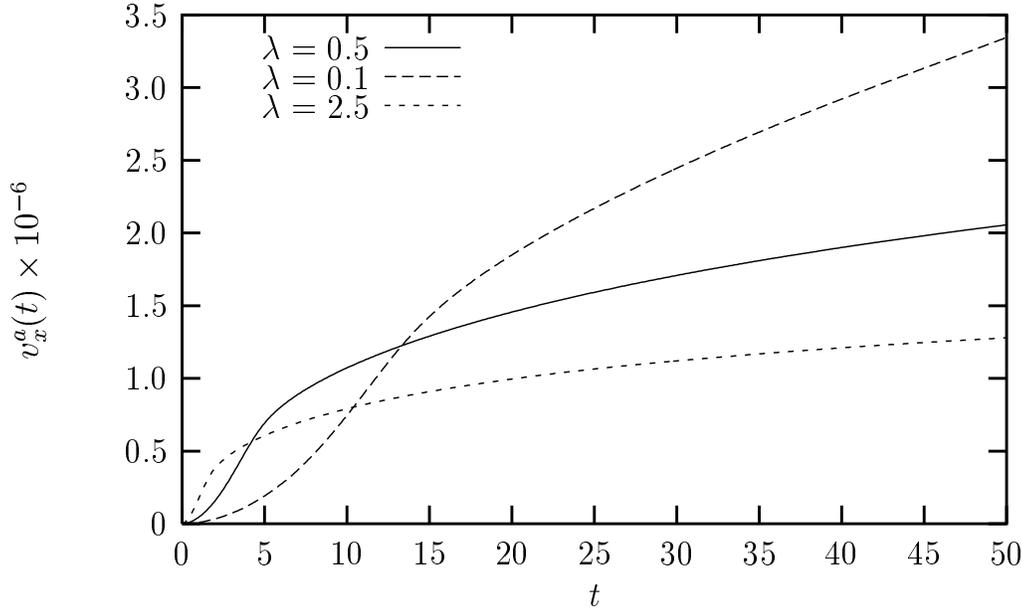}}
     \caption{Maximal value of the generated
fast magnetoacoustic wave, $V_x^a(t)$, versus time.
Solid, long dashed and short dashed curves represent
solutions of Eq.(12) using 2D MacCormak code on
$4000 \times 4000$ spatial resolution for $\lambda=0.5,0.1,2.5$ 
accordingly.}
\label{fig11}
\end{figure}
It is worthwhile to note the 1D results (Fig. 9)
of the variation of $\lambda$-parameter show a similar
trend, i.e. start of the saturation process
earlier for larger $\lambda$'s. However, as in the
1D results of Botha et al. 2000 (see p. 1191 in their
paper for the discussion) the scaling of the saturation
amplitudes with $\lambda$ is not the same as in 2D or 
full MHD simulations.

In Fig. 12 we investigate the scaling of saturation levels
of the generated fast magnetoacoustic waves with the
$\delta$-parameter, which controls width of the
phase-mixed  Alfv\'en pulse over $z$-coordinate.
Fig. 12 confirms that with the decrease of $\delta$, i.e.
for narrower Gaussian pulses, saturation
levels of the generated fast magnetoacoustic waves
increase and vice versa.
\begin{figure}
     \resizebox{\hsize}{!}{\includegraphics{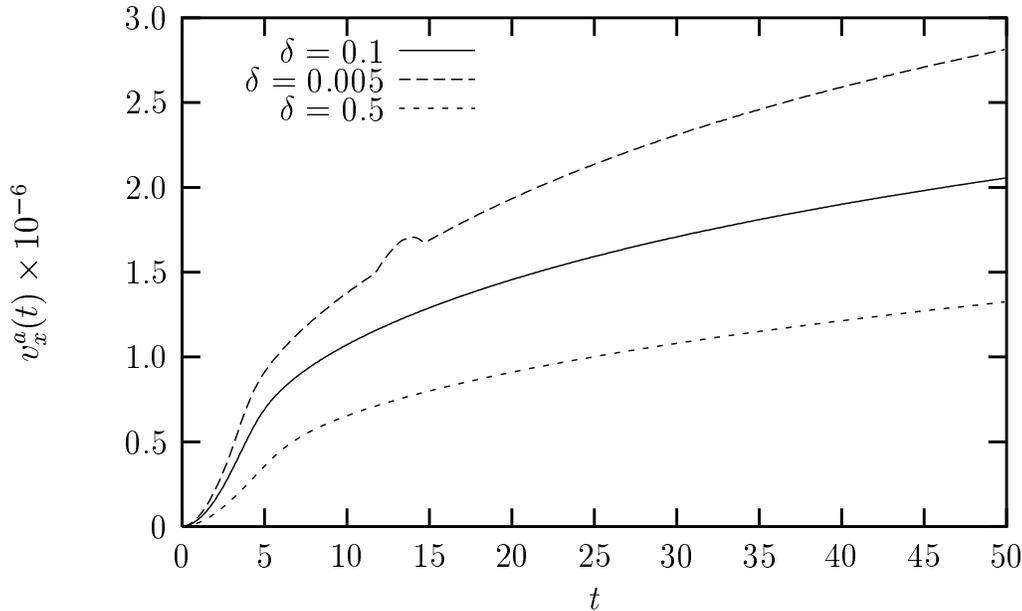}}
     \caption{Maximal value of the generated
fast magnetoacoustic wave, $V_x^a(t)$, versus time.
Solid, long dashed and short dashed curves represent
solutions of Eq.(12) using 2D MacCormak code on
$4000 \times 4000$ spatial resolution for $\delta=0.1,0.005,0.5$ 
accordingly.}
\label{fig12}
\end{figure}

Finally, we would like to comment on the choice of
the upper bound of our simulation time, which we have
set to $t=50.0$. DeMoortel et al. 1999 have found
that in linear phase mixing (with the dissipation
effects included) shear viscosity dissipation reaches its
maximum when
the harmonic Alfv\'en wave is 10 wavelengths out of phase.
Assuming that this result holds for the traveling
Gaussian Alfv\'en pulse, we can estimate time the $t_*$ it takes
for the classic shear viscosity to become significant.
The distance between the wings of the distorted
phase mixed Alfv\'en pulse (see Fig. 4) is
$(C_A(+\infty)- C_A(-\infty))\,t_*$. If we use the
pulse width, $2 \sqrt{\delta}$, instead of wavelength
in the original DeMoortel et al. 1999's estimate,
we can write $(C_A(+\infty)- C_A(-\infty))\,t_*=
10 \times (2 \sqrt{\delta})$. Inserting
our values for the asymptotic Alfv\'en speeds
and $\delta=0.1$ in the latter equality, we deduce
$t_* \approx 11$. Therefore, there is little
point running the present model for $t>t_*$, because 
as shown by DeMoortel et al. 1999, viscosity would
be significant for such later times and
this is excluded from our current model.

\section{Discussion}

The motivation of this paper was three-fold.
Botha et al. 2000 concluded that
for harmonic phase-mixed Alfv\'en perturbations the
generated fast magnetoacoustic waves do not grow
to a substantial fraction of the
Alfv\'en wave amplitude.
Firstly, we wanted to test whether a single phase-mixed 
Gaussian Alfv\'en pulse behaves in the same way.
Secondly, to test the validity of the weakly non-linear
analytic description of Nakariakov, Roberts \& Murawski 1997
for large times.
Thirdly, to check whether the rapid growth (in time)
of the generated fast magnetoacoustic waves still holds
at the later stages.

Our present analysis has clearly demonstrated that:
\begin{itemize}
\item Small-amplitude, phase-mixed,  single
 Alfv\'en pulses do not
generate fast magnetoacoustic waves with large amplitudes.
The saturation is due to the destructive wave interference.
\item We have confirmed that the weakly non-linear
analytic description of Nakariakov, Roberts \& Murawski 1997
(Eq.(12))
is valid and produces results that are entirely
consistent with the simulation results from fully
non-linear MHD equations (1)-(4) {\it at all stages}, provided the
initial amplitude of the Alfv\'en perturbation is small.
\item We have found that the  fast magnetoacoustic wave, generated by the
phase mixed single Alfv\'en  pulse, at the later stages
grows much slower than at earlier times 
due to the destructive wave interference.
\end{itemize}

We have also established that the
stronger transverse gradients (larger
$\lambda$'s) cause earlier start of the fast magnetoacoustic waves
saturation process, and in turn, yield lower saturation levels. 
Whereas, weaker gradients (smaller
$\lambda$'s) result in later onset of the saturation, and in turn,
higher saturation levels.
This result,  actually, validates our explanation 
of the fast magnetoacoustic wave saturation by the
destructive wave interference effect. Namely, larger $\lambda$
means steeper transverse gradient, i.e. narrower driver,
acting from the {\it beginning}. Therefore, destructive wave interference
effect starts to a play role earlier yielding faster shut down
in the growth of $V_x$. For the the weaker gradients, i.e.
small $\lambda$ exactly the opposite is true.
This result is in agreements with the one
obtained by Botha et al. 2000 for a harmonic 
Alfv\'en wave (see Fig. 3 in
Botha et al. 2000 for comparison).
 It is worthwhile to mention, however, that on one hand,
contrary to the one's
expectations, steeper inhomogenities of the Alfv\'en speed do not
provide the higher saturation levels
of the fast wave generation. 
In other words, with increase of $\lambda$ saturation levels
decrease.
On the other hand, in the homogeneous plasma ($\lambda=0$)
there is no
generation of the fast magnetoacoustic wave by a plane
(with respect to $x$-coordinate) Alfv\'en wave at all, i.e. 
$V_x^a(t) \to 0$ when $\lambda \to 0$.
Therefore, 
there is a certain optimal value of gradient of the
inhomogeneity for which the fast wave saturation level is maximal.
At this stage we have not performed full parametric study in order to
determine the optimal value of $\lambda$ due to 
the vast requirements for CPU time.

The mechanism of indirect plasma heating by the
fast magnetoacoustic waves that are non-linearly generated
by the phase-mixed  Alfv\'en perturbation, discussed
in this paper, is certainly applicable for the
coronal heating problem through dissipation of
fast magnetoacoustic waves in open (e.g. plumes)
and closed (loops) coronal plasma structures.
Although, we found that the proposed mechanism,
in the considered set up, 
is limited by destructive wave interference,
further work is needed to find possible ways
around of the encountered difficulty.
Possible cures might be considering
2D density inhomogeneity profiles,
that would prevent the driver of Eq.(12) from
becoming too narrow, and/or inclusion of the dissipation into our model,
that would prevent the transverse scale collapse, which all
ultimately will lead to the decrease of the
destructive wave interference.

\acknowledgements
Numerical calculations of this work were
done using the PPARC funded Compaq MHD Cluster in St. Andrews.

\end{document}